\documentclass[astra]{copernicus}

\begin{document}

\title{POLAR: A Space-borne X-Ray Polarimeter for Transient Sources}

\author{Silvio Orsi on behalf of the POLAR collaboration}

\affil{DPNC, 24 Quai Ernest-Ansermet, Universit\'e de Gen\`eve, 1205 Geneva, Switzerland}

\runningtitle{The X-Ray Polarimeter POLAR}

\runningauthor{S.~Orsi}

\correspondence{S.~Orsi\\ (Silvio.Orsi@unige.ch)}

\received{}
\pubdiscuss{} 
\revised{}
\accepted{}
\published{}


\firstpage{1}

\maketitle

\begin{abstract}
POLAR is a novel compact Compton X-ray polarimeter designed to measure the linear polarization of the prompt emission of Gamma Ray Bursts (GRB) and other strong transient sources such as soft gamma repeaters and solar flares in the energy range 50$-$500~keV. 
A detailed measurement of the polarization from astrophysical sources will lead to a better understanding of the source geometry and emission mechanisms. 
POLAR is expected to observe every year several GRBs with a minimum detectable polarization smaller than 10\%, thanks to its large modulation factor, effective area, and field of view.
POLAR consists of 1600 low-Z plastic scintillator bars, divided in 25 independent modular units, each read out by one flat-panel multi-anode photomultiplier. 
The design of POLAR is reviewed, and results of tests of one modular unit of the engineering and qualification model (EQM) of POLAR with synchrotron radiation are presented. 
After construction and testing of the full EQM, we will start building the flight model in 2011, in view of the launch foreseen in 2013.
\end{abstract}

\introduction
Gamma Ray Bursts are cosmic explosions that happen at cosmological distances at random times in random places in the Universe. 
Polarization measurements of GRB prompt emission may distinguish between the proposed theoretical models: synchrotron with random field (the fireball model)~\citep{piran2004}, synchrotron with ordered field (the electromagnetic model)~\citep{lyutikov2003} and Compton drag (cannonball model)~\citep{dar2004}.
All of them relate the emission of the GRB to the creation of a black hole, differing in the physical processes involved in the  $\gamma$--ray generation, and also in the level of linear polarization of the emitted photons. 
The direction and the level of polarization of high-energy photons emitted by astrophysics sources such as GRBs are therefore very good observable candidates for the understanding of the corresponding emission mechanisms, source geometry and strength of magnetic fields at work~\citep{lazzati2006, toma2008}.
In view of the power and the lack of precise polarization studies of GRBs, s
Several X and $\gamma$--ray polarimeters have been proposed and are under development to perform precise polarization studies of GRBs. 
Some examples are POLAR~\citep{produit2005}, GRAPE (Gamma-Ray Polarimeter Experiment~\citep{bloser2006}) POET (Polarimeters for Energetic Transients~\citep{hill2008}), CIPHER (Coded Imager and Polarimeter for High Energy Radiation~\citep{curado2003}), PHENEX (Polarimetry for High ENErgy X rays~\citep{gunji2007}), XPOL~\citep{costa2007} and POLARIX~\citep{costa2006}. 
Furthermore, polarimeters designed for studying fixed sources, like PoGOLite~\citep{kamae2008}, are also able to measure the polarization of GRBs if they happen to appear in their field of view.

\begin{figure}[b]
\vspace{2mm}
\begin{center}
\includegraphics[width=8.3cm]{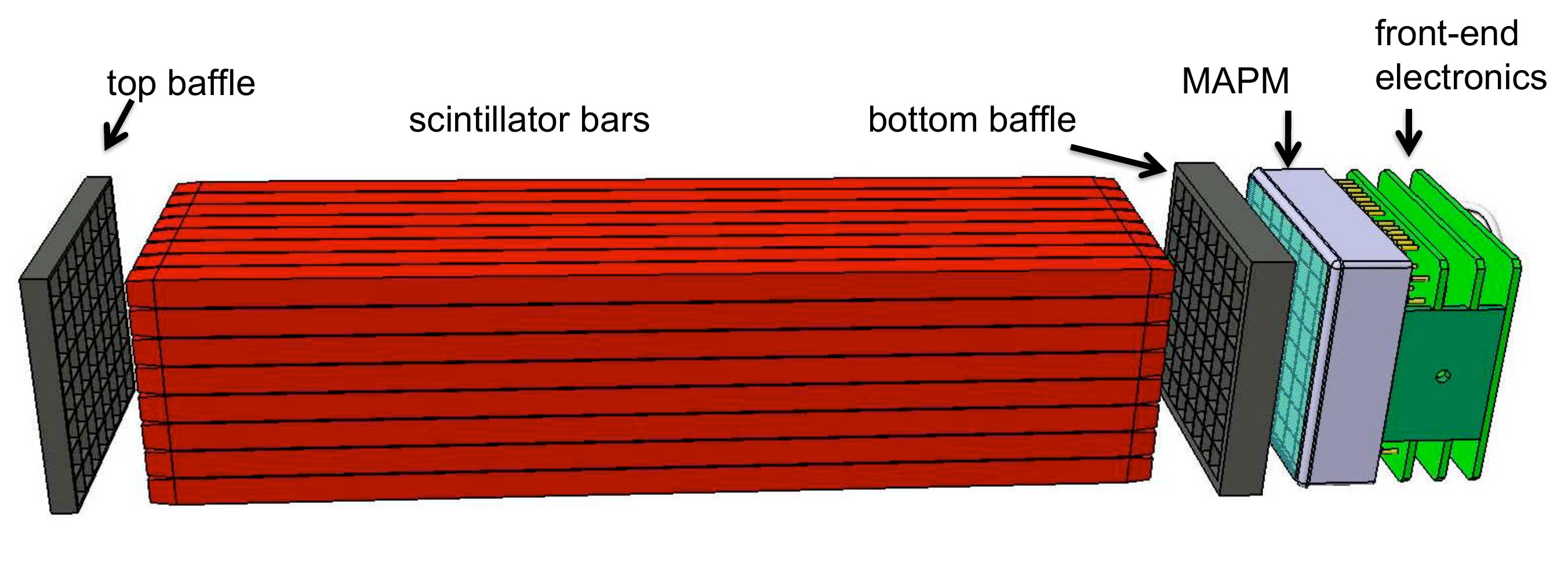}
\end{center}
\caption{Exploded view of one POLAR modular unit without the carbon fiber socket.}
\label{fig:modulecad}
\end{figure}

\begin{figure}[t]
\vspace*{2mm}
\begin{center}
\includegraphics[width=7cm]{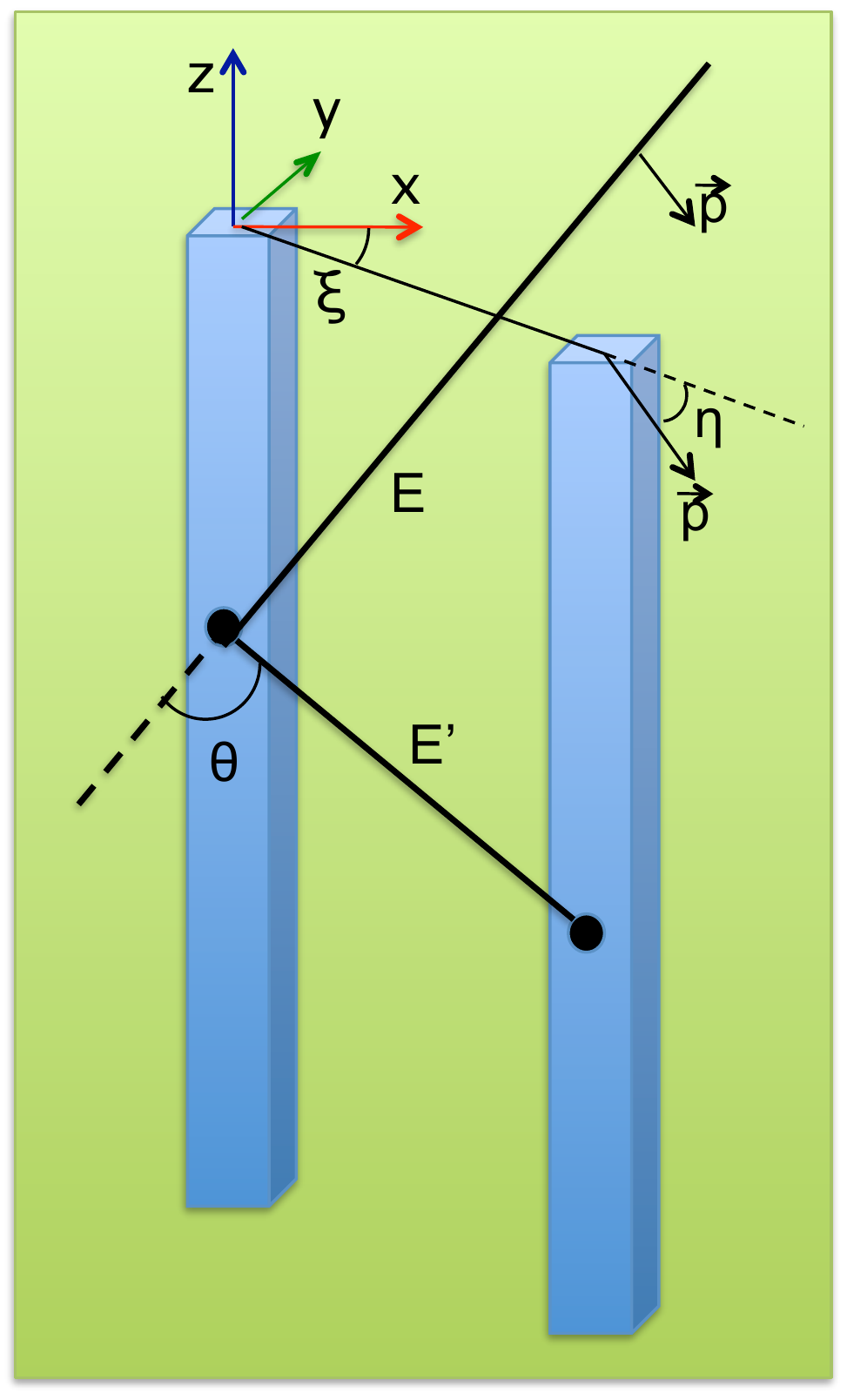}
\end{center}
\caption{Definition of the coordinate system and of angles for an incoming photon interacting in 2 bars.}
\label{fig:angles}
\end{figure}

\section{The POLAR Detector}
\label{sec:polar}
POLAR~\citep{produit2005} is a novel compact detector aimed at the polarization measurement of GRBs in the energy range 50$-$500~keV.
It consists of 40$\times$40 plastic scintillator bars (each 6$\times$6$\times$200mm$^3$), organized in 25~independent modular units, with 64 bars each.
Each unit (figure~\ref{fig:modulecad}) is read-out by a flat panel multi anode photomultiplier tube (MAPMT; H8500, Hamamatsu), mechanically coupled to the bottom of the scintillator bars via a thin optical pad, and enclosed in a 1~mm carbon fiber socket.
A plastic baffle placed between the bars and the MAPMT increases the mechanical stability and reduces the crosstalk between channels below 10\%.
The electrical signals coming from the MAPMT are first processed by ASICs and FPGAs at the front-end electronics, then sent to the POLAR central computer, where the trigger decision is taken considering the outputs of all modular units.
This modular design provides a good mechanical stability and facilitates the interchange of modules during the testing phase of the detector.
POLAR is characterized by 
a relatively large effective area ($\sim$80~cm$^2$), large field of view ($\sim$1/3~of the sky) and large modulation factor (30$-$40\%) for GRB measurements.
A flight opportunity for POLAR on the future Chinese Tian-Gong Space Station is currently under consideration.

POLAR is based on Compton scattering, by which photons tend to scatter at an angle perpendicular to the photon polarization vector, according to the Klein-Nishina formula.
The incoming photon undergoes Compton scattering in one bar; if the scattered photon interacts in at least one other bar (via Compton or photoelectric effect), we accept the event as `valid'. 
The interaction locations of the 2 bars with highest energy depositions 
define the azimuthal scattering angle $\xi$ (see figure~\ref{fig:angles}).
The distribution of $\xi$ for all photons from a transient source is referred to as modulation curve and is described by the function: 
\begin{equation}
f(\xi) = K \cdot \{1 + \mu \cos [2 ( \xi - \xi_0 ) + \pi ] \}, 
\label{eq:modcurve}
\end{equation}
where the angle of polarization $\xi_0$ and the modulation factor $\mu$ are obtained from a fit to the data.
For each event the angle $\xi$ is obtained by measuring the angle between the 2 bars with highest energy depositions, excluding neighbor channels, whose signal is dominated by crosstalk. 
The degree of linear polarization is $\Pi=\mu/\mu_{100}$, where $\mu_{100}$ is the modulation factor for a 100\% polarized photon beam, obtained with Monte Carlo simulations and verified with beam test data. 
A sinusoidal term with period $\pi/2$ adds to the function due to the square geometry of the POLAR instrument.

Extensive Monte Carlo simulations to study the response of POLAR to GRB with different spectral characteristics and placed in several locations in the sky with respect to POLAR have been performed and show very good polarimetric performance~\citep{suarez2010phd}.
These simulations show also that GRB photons scattered in the mechanical structure of the spacecraft into POLAR have a rather small effect: they reduce the $\mu_{100}$ from $\sim$30\% to $\sim$26\% for a typical GRB.

\begin{figure}[t]
\vspace*{2mm}
\begin{center}
\includegraphics[width=8.3cm]{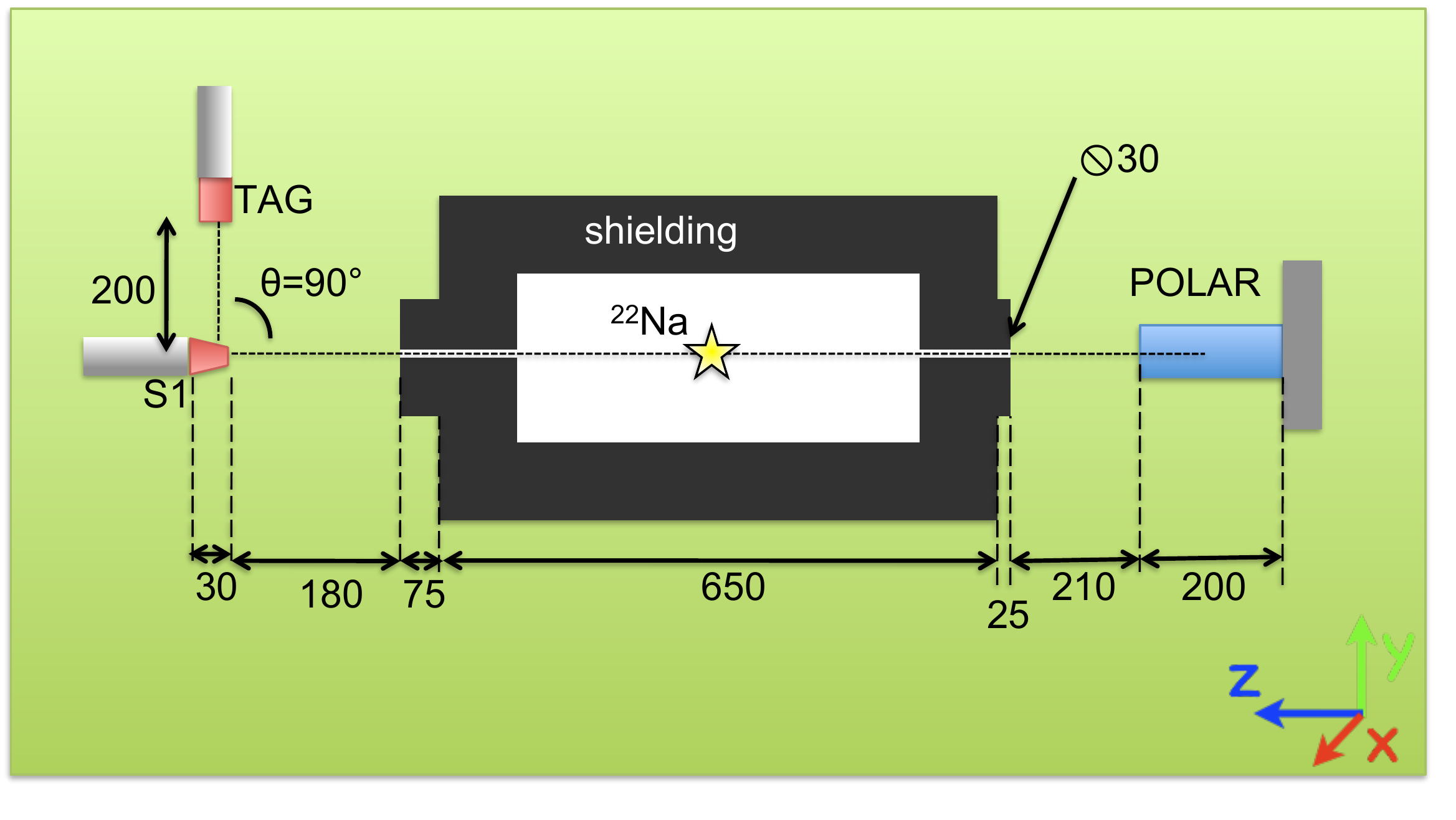}
\end{center}
\caption{Scheme of the experimental setup for measurements with a polarized source of 511~keV photons.}
\label{fig:testbenchsetup}
\end{figure}

\begin{figure}[t]
\vspace*{2mm}
\begin{center}
\includegraphics[width=8.3cm]{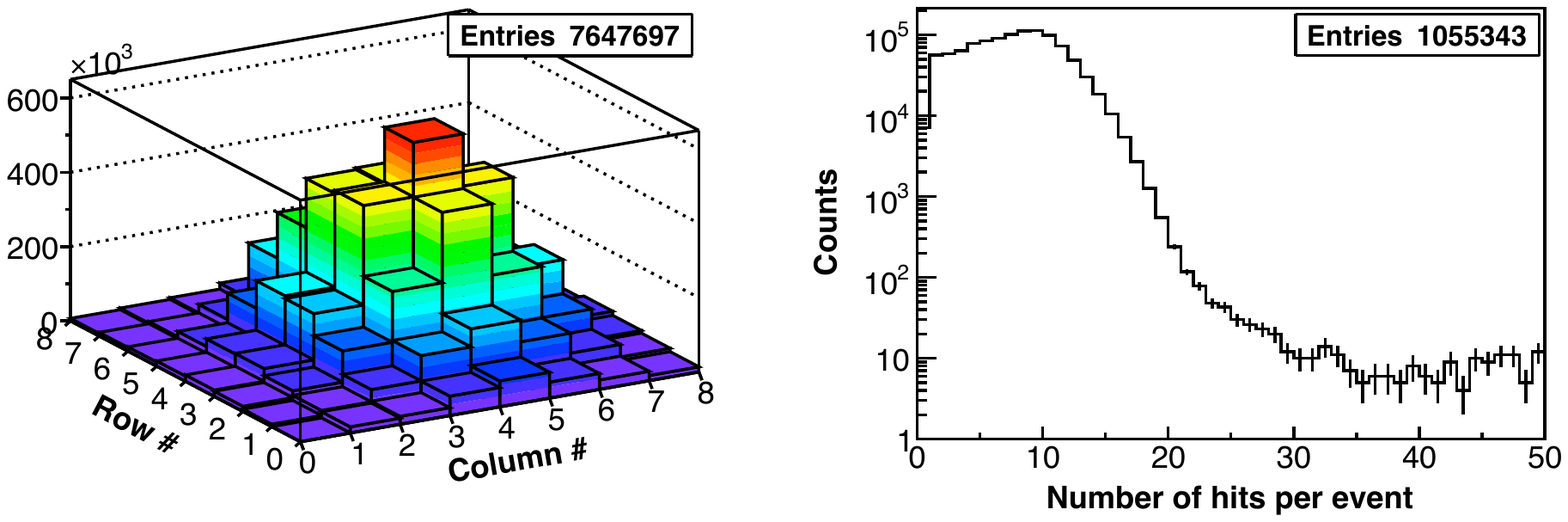}
\end{center}
\caption{Beam position for the measurements with a Na-22 source.}
\label{fig:lablego}
\end{figure}

\begin{figure}[t]
\vspace*{2mm}
\begin{center}
\includegraphics[width=8.3cm]{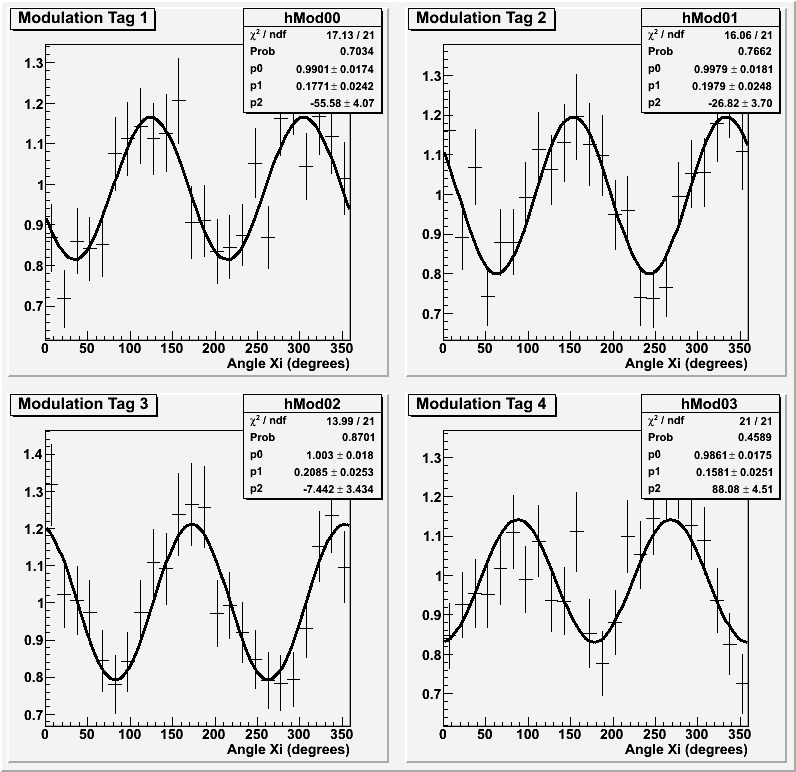}
\end{center}
\caption{Modulation curves obtained with photons from the Na-22 source, tagged respectively at $\xi \sim$0$^{\circ}$, 30$^{\circ}$, 90$^{\circ}$ and 180$^{\circ}$.}
\label{fig:labmod}
\end{figure}

\begin{figure}[th]
\vspace*{2mm}
\begin{center}
\includegraphics[width=8.3cm]{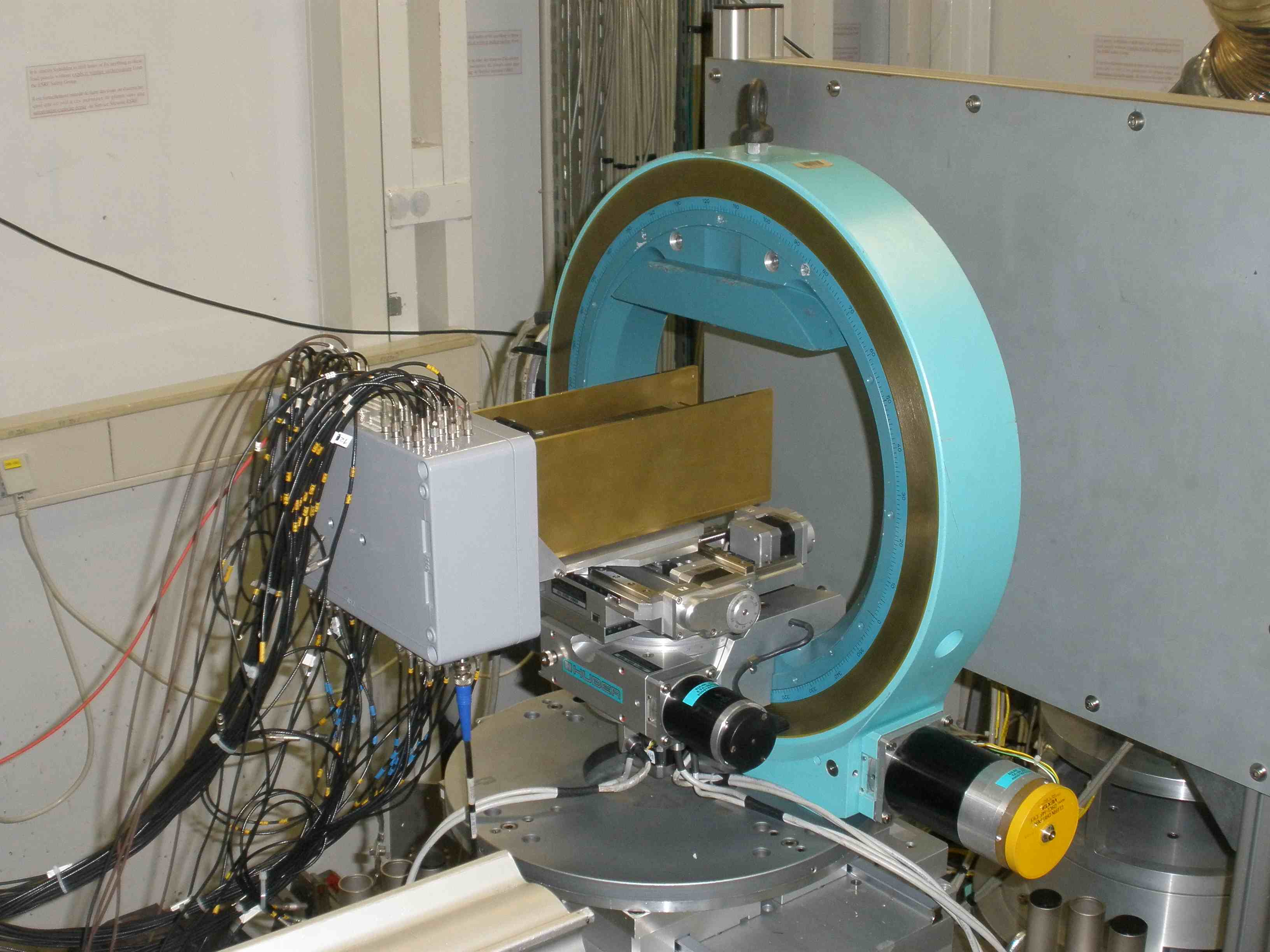}
\end{center}
\caption{Experimental setup for the measurements at ESRF.}
\label{fig:esrfsetup}
\end{figure}

\begin{figure}[t]
\vspace*{2mm}
\begin{center}
\includegraphics[width=8.3cm]{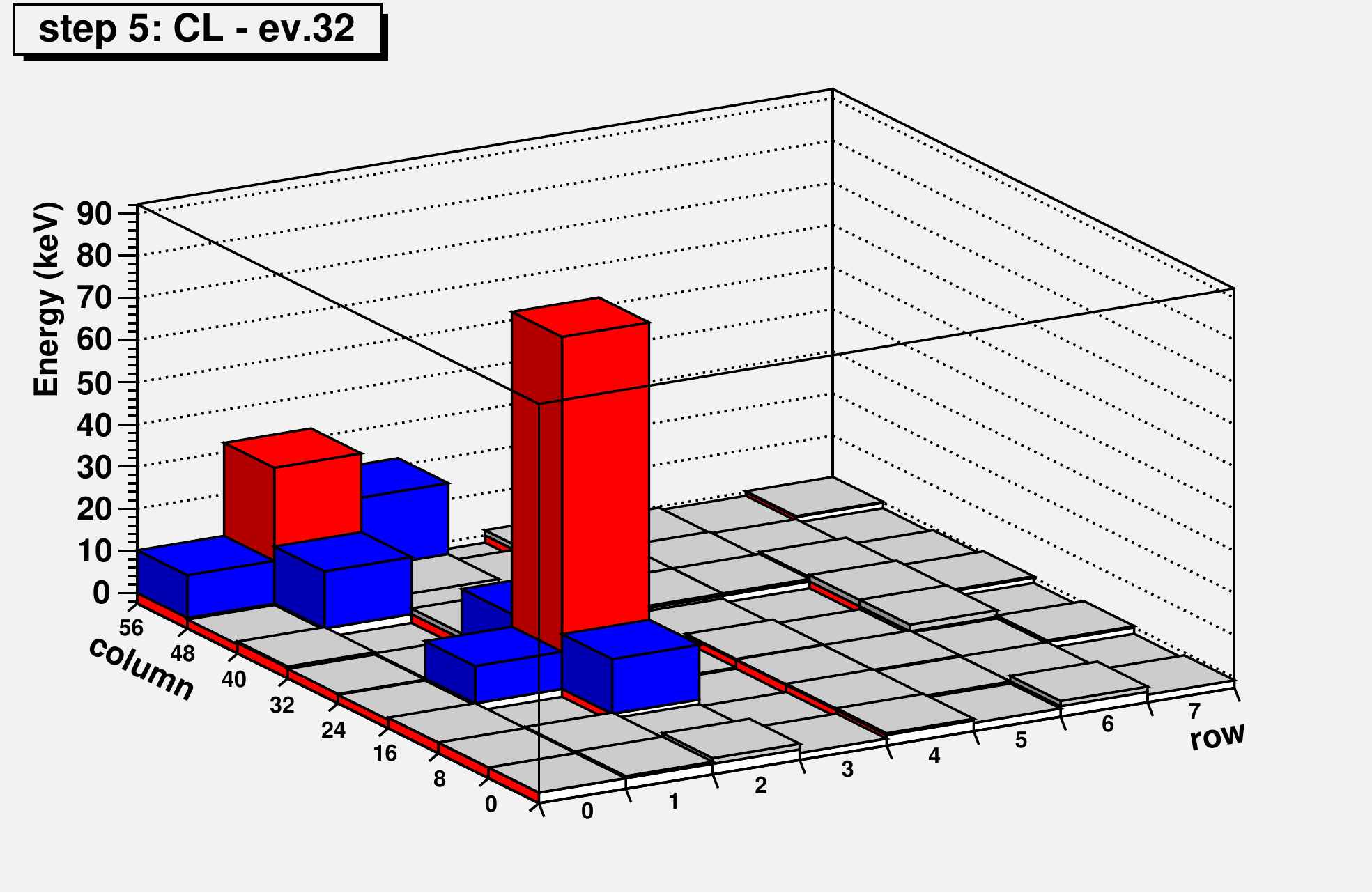}
\end{center}
\caption{Typical event. The 2 red bars are the 2 selected hits, while the blue bars are the other channels with deposited E$>$5~keV.}
\label{fig:esrfevent}
\end{figure}

\begin{figure*}[tb]
\vspace*{2mm}
\begin{center}
\includegraphics[width=8.3cm]{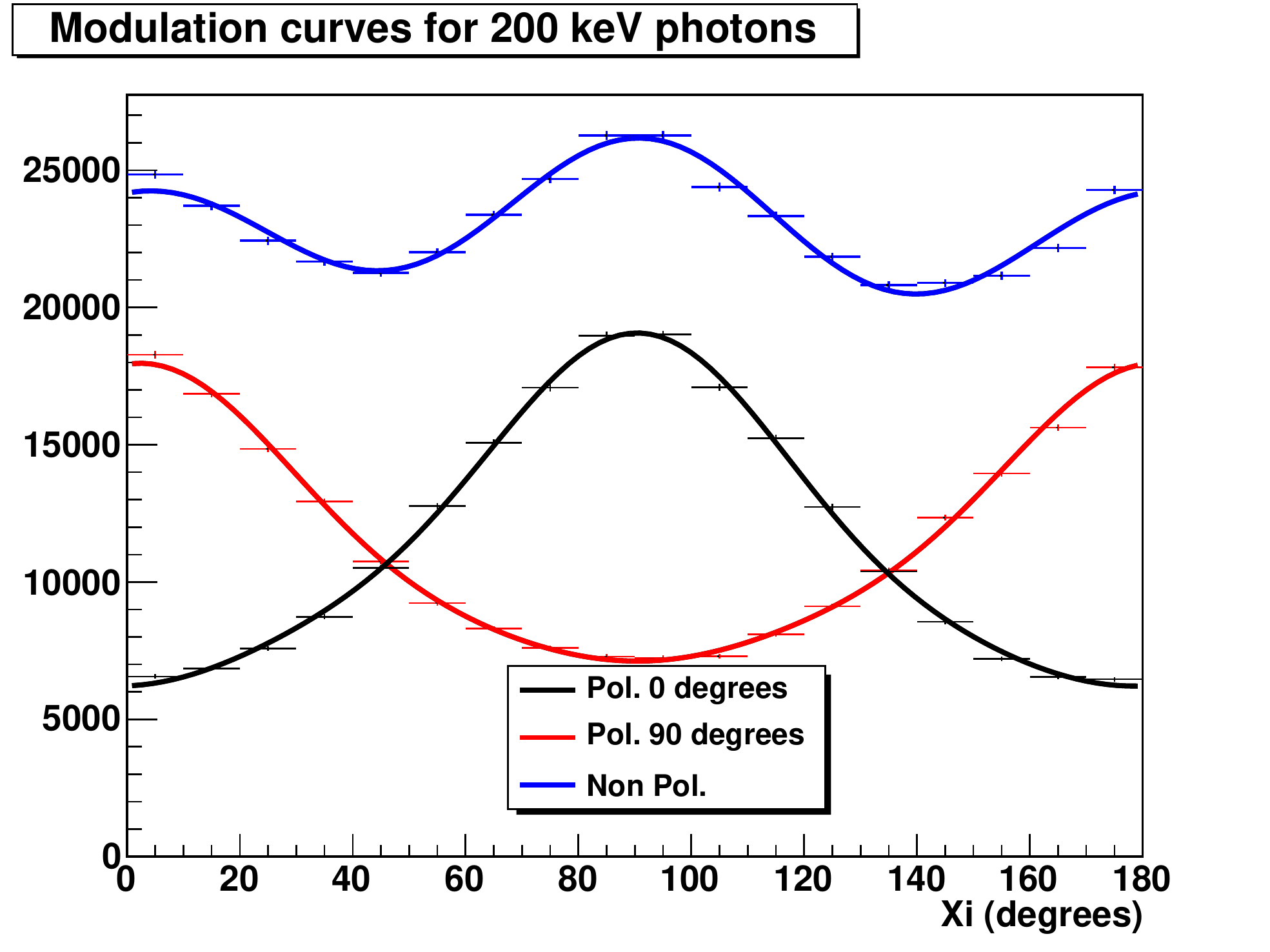}
\includegraphics[width=8.3cm]{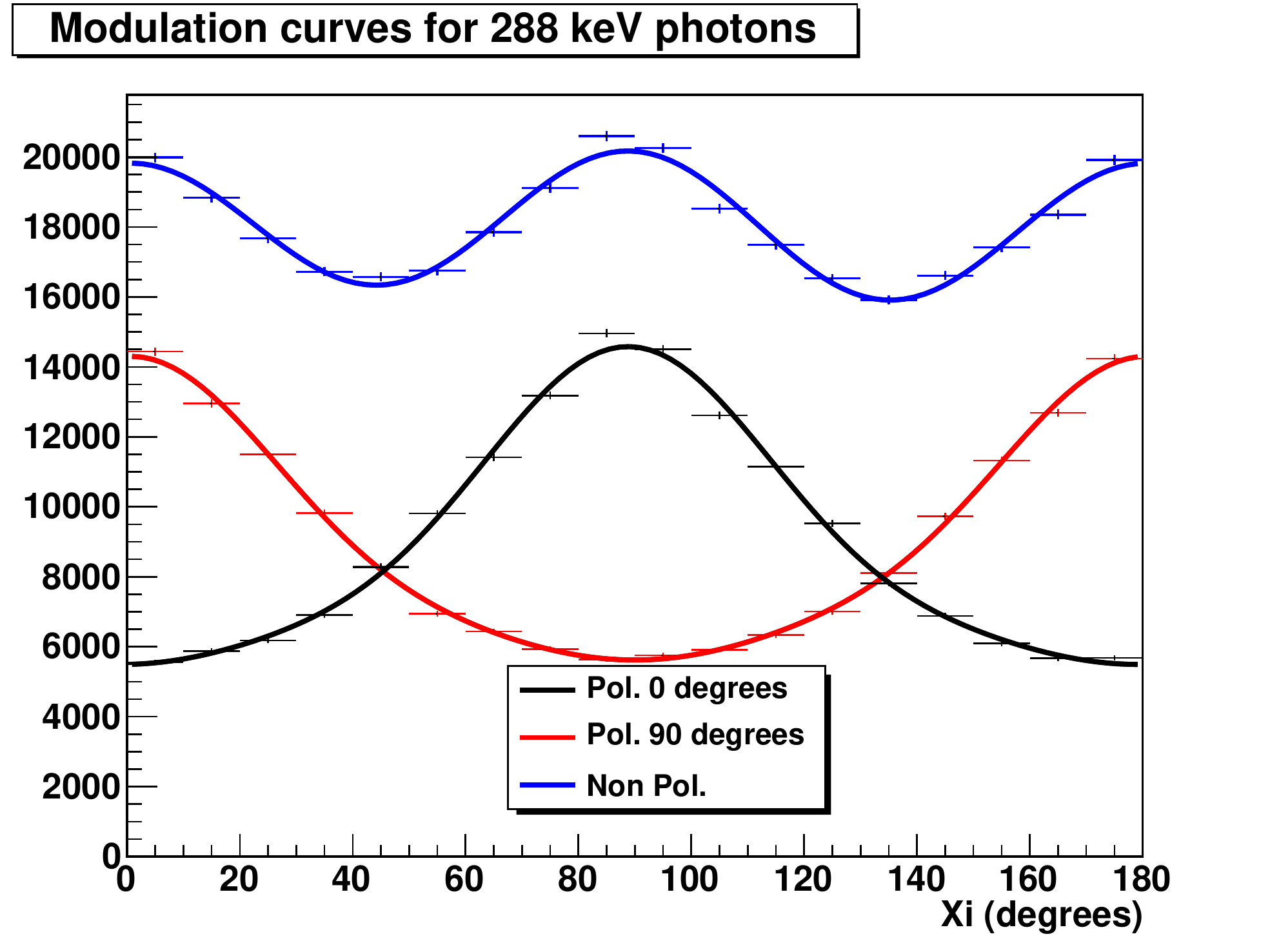}
\includegraphics[width=8.3cm]{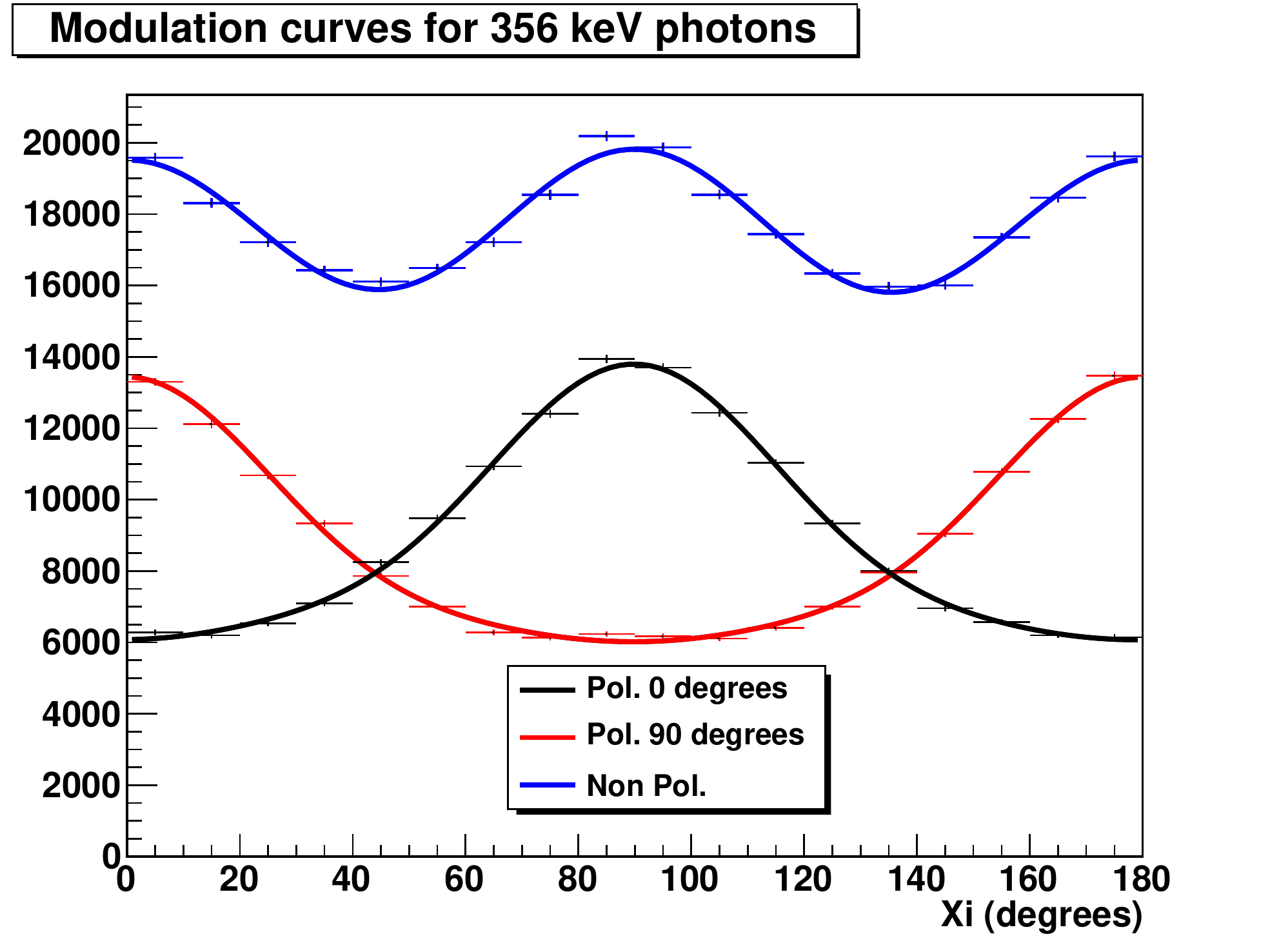}
\includegraphics[width=8.3cm]{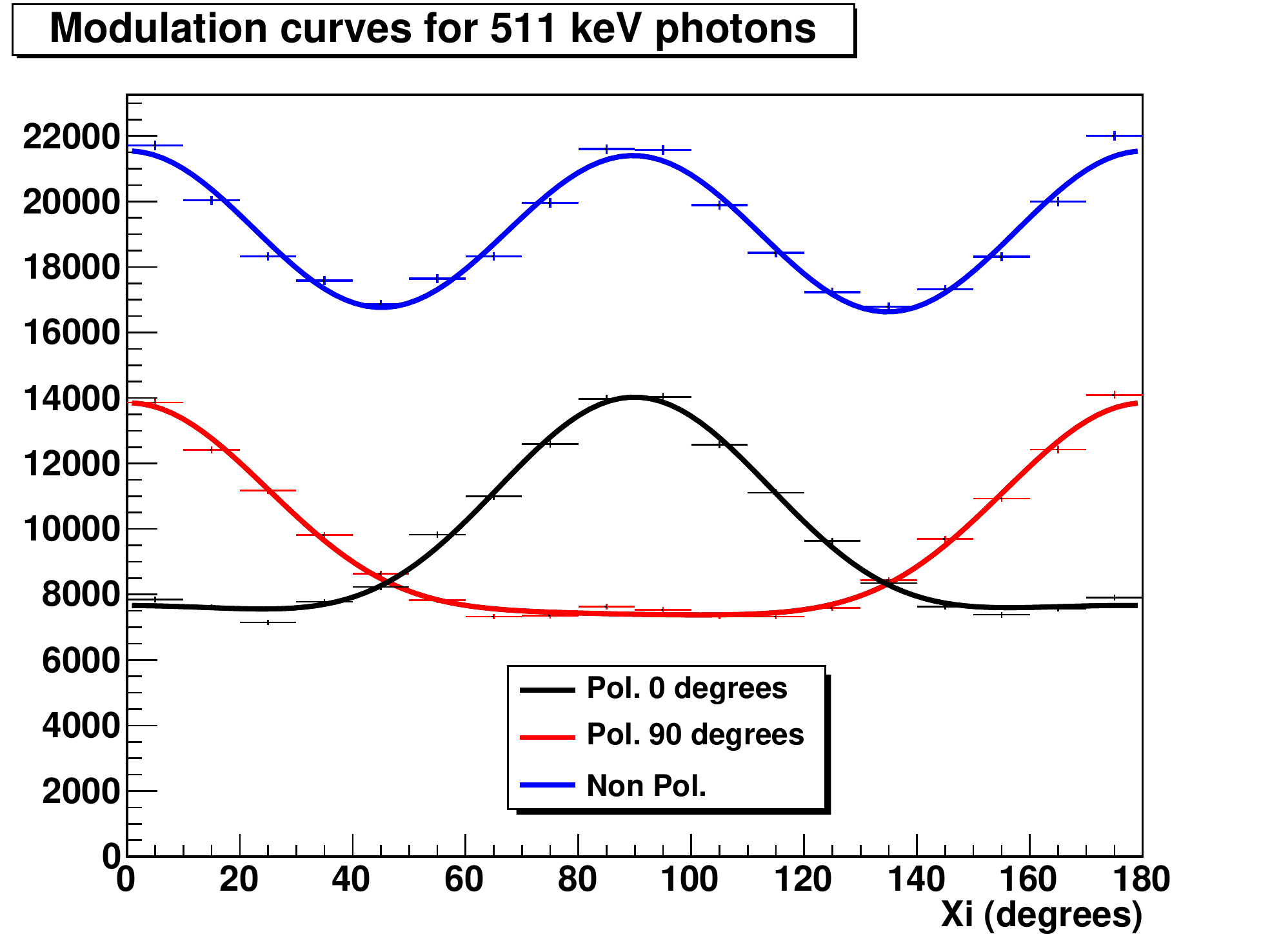}
\end{center}
\caption{Measured modulation curves for input beams of 200, 288, 356 and 511 keV for polarized and non-polarized photons (analysis threshold on both hits: 30~keV). The modulation amplitude due to the square geometry is $\sim$10\%.}
\label{fig:esrf4modcurves}
\end{figure*}

\section{Measurements with a polarized radioactive source}
\label{sec:na22}
We have set up an experimental test bench with a partially polarized radioactive source at the University of Geneva.
At the center of the setup, shown in figure~\ref{fig:testbenchsetup}, a Na-22 radioactive source emits positrons, which are captured by the surrounding matter to form positronium atoms. 
These decay 
into two correlated photons of 511~keV, emitted back-to-back.
A fraction of the photon pairs will exit through 2 circular holes and hit the POLAR modular unit under test (right side of figure~\ref{fig:testbenchsetup}) and a plastic scintillator (S1, shown on the left).
Four identical NaI detectors  (TAG) are placed around S1 at four azimuth scattering angles $\xi$, to tag photons that Compton scatter in S1 by $\theta \sim 90^{\circ}$ and $\xi$ respectively $\sim$0$^{\circ}$, 30$^{\circ}$, 90$^{\circ}$ and 180$^{\circ}$.
The correlated photon 
emitted in the opposite direction is detected in POLAR, as previously schematically shown in figure~\ref{fig:angles}.
All events characterized by coincident energy depositions in both S1 and in the POLAR unit are recorded on disk.
Figure~\ref{fig:lablego} shows the beam position, i.e.~the number of hits detected in each of the 64 channels.
As expected by geometry, most of the hits are recorded in the central bars, where the photon beam (beam size $\sim$1~cm) enters the modular unit.
Due to the finite solid angle subtended by the TAG detectors, there is a systematic uncertainty on the polarization direction of $\sim$2$^{\circ}$.
The analysis on the data is performed considering the `modulation ratio', i.e. the ratio of the modulation curves for polarized and unpolarized data.
This procedure suppresses geometry effects and removes residual crosstalk between channels, in large part already reduced since hits in neighbor bars are neglected.
The four modulation curves obtained with this setup are shown in figure~\ref{fig:labmod}; the low statistics reflects the fact that the rate of coincidences in S1  and POLAR is small ($<$10~Hz), and that only $\sim$1\% of the events present activity in a TAG detector.
Monte Carlo simulations show that  $\mu_{100} = 31.3 \pm 0.7\%$ for photons of 511~keV entering one modular unit from the zenith.
This allows us, together with the experimental measurement of $\mu$ ($17 \pm 2\%$), to reconstruct the polarization of the four data samples, $\Pi \sim 56 \pm 4\%$, in agreement with simulations of the experimental setup.

\section{Measurements with synchrotron radiation}
\begin{figure}[t]
\vspace*{2mm}
\begin{center}
\includegraphics[width=8.3cm]{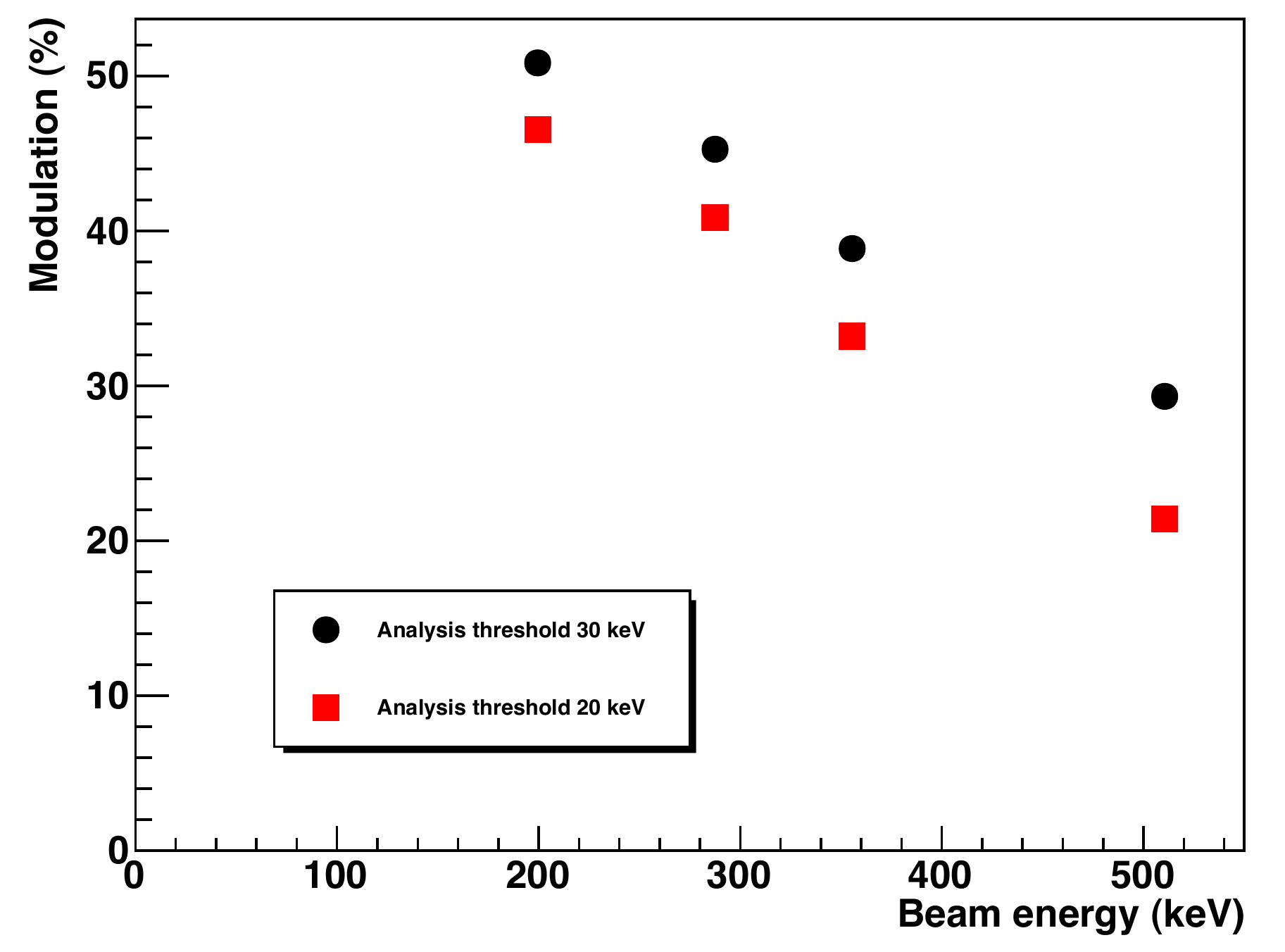}
\end{center}
\caption{Modulation factor dependence on the beam energy.}
\label{fig:esrfmodvsthr}
\end{figure}

In December 2009 we performed a systematic calibration of the polarimetric response of one modular unit of POLAR with high-energy 100\% polarized photons at the European Synchrotron Radiation Facility (ESRF) in Grenoble~\citep{orsi2010, suarez2010phd}.
The device under test (figure~\ref{fig:esrfsetup}) has dimensions $\sim5 \times 5 \times20$~cm$^3$, weighs $\sim$10~kg including mechanical support and shielding, and is placed in an Eulerian cradle, mounted on a large moving table.
The experiment consists of recording all pairs of bars that show a coincident energy deposition.
Figure~\ref{fig:esrfevent} shows a typical event, where 8 channels have deposited energy larger than 5~keV, and the red channels are the 2 hits selected to calculate $\xi$.
A photon beam of size $0.5 \times 0.5$~mm$^2$ and parallel to the detector axis is directed into the center of each bar, one after the other, and provides a uniform illumination of the detector with 1.28 $\cdot 10^6$ triggers ($2 \cdot 10^4$ events per bar).
From the distribution of the azimuthal scattering angle $\xi$ we reconstruct the angle of polarization $\xi_0$ of the incoming photon beam and the modulation factor $\mu_{100}$, i.e.~the response of the instrument to a fully polarized photon flux.
About 40 million of triggers have been collected for beam energies of 200, 288, 356 and 511~keV.
The results demonstrate the excellent polarimetric capabilities of the instrument.
Figure~\ref{fig:esrf4modcurves} shows the modulation curves obtained for photon beams with horizontal (black line) and vertical (red) polarization and also for non-polarized photons (blue), where the minimum energy deposited per bar is 30~keV.
The non-polarized photons are obtained as the sum of samples with orthogonal polarization, and their distribution is described by a sinusoidal function with period $\pi/2$ and amplitude $\sim$10\% due to the square geometry of POLAR.
The results obtained 
are in agreement with Monte Carlo simulations~\citep{orsi2010}.
Fits to these curves lead to modulations $\mu_{100}$ between $\sim$30 and 50\%, as shown in figure~\ref{fig:esrfmodvsthr} (black circles); the red square markers indicate the reconstructed $\mu_{100}$ in case a minimum energy of 20~keV per channels is used.
With increasing threshold on the deposited energy larger scattering angles $\theta$ are selected, which leads to larger modulation factors, at the expense of reducing the statistics.

\conclusions[Conclusions and Outlook]
POLAR is a Compton polarimeter designed to measure the polarization of the prompt emission of GRBs.
One modular unit of POLAR has been tested in the laboratory with polarized photons of 511~keV from a radioactive source and with 100\% polarized synchrotron radiation of 200$-$511~keV.
Both tests have shown excellent performance in measurements of polarization. 
The polarimetric response of a larger POLAR instrument with synchrotron radiation in the range 50$-$500~keV is planned for the next year to verify the localization method of GRB, currently based on simulations~\citep{suarez2010}.
The knowledge of the GRB position in the sky within few degrees is needed since the modulation factor $\mu_{100}$ depends on the relative position of the GRB with respect to POLAR.
A flight onboard the Chinese Space Station in $\sim$2013 is under consideration. 

\begin{acknowledgements}
The authors acknowledge the European Synchrotron Radiation Facility for provision of synchrotron radiation facilities and thank Veijo Honkimaki, responsible of the beam line ID15 at ESRF, for his continuous help and support during all the phases of the test.

\end{acknowledgements}

\end{document}